\documentclass[twocolumn, final]{article}          
\usepackage[a4paper,
            bindingoffset=0.5cm,
            left=1.25cm,
            right=1.25cm,
            top=1.5cm,
            bottom=1.5cm,
            footskip=1cm]{geometry}
\usepackage[utf8]{inputenc}
\usepackage[T1]{fontenc}
\usepackage{graphicx, mathptmx, newtxtext, newtxmath} 
\usepackage{subcaption, rotating, adjustbox, colortbl, booktabs, siunitx, xargs, xpatch, float, tabularray, capt-of, authblk, bm, xcolor, amsmath} 
\usepackage{threeparttable}
\usepackage[switch]{lineno}
\usepackage[acronym]{glossaries}
\usepackage[]{natbib}
\UseTblrLibrary{booktabs} 

\newacronym{VR}{VR}{Virtual Reality}
\newacronym{Dpt}{D}{Diopter or dioptre}
\newacronym{IPD}{IPD}{Inter- Pupillary Distance}
\newacronym{fov}{FoV}{field of view}
\newacronym{hmd}{HMD}{head-mounted display}
\newacronym{pals}{PALs}{progressive addition lenses}
\newacronym{Pals}{PALs}{Progressive addition lenses}
\newacronym{vr}{VR}{virtual reality}
\newacronym{gbr}{GBR}{gradient boosting regressor}
\newacronym{UI}{UI}{user interface}
\DeclareSIUnit\diopter{D}
\DeclareSIUnit\px{px}

\usepackage[hyphens]{url}
\usepackage[]{hyperref}
\hypersetup{breaklinks=true}
\begin{document}
\title{\acrfull{VR} as a testing bench for consumer optical solutions: A machine learning approach (\acrshort{gbr}) to visual comfort under simulated \acrfull{pals} distortions}

\author[1]{Miguel García García}
\author[1]{Yannick Sauer }
\author[2]{Tamara Watson }
\author[1,3]{Siegfried Wahl }
\affil[1]{Institute for Ophthalmic Research, University of Tuebingen, Elfriede-Aulhorn-Straße 7, Tübingen, 72072 }
\affil[2]{School of Social Science and Psychology, Western Sydney University, New South Wales, Australia }
\affil[3]{Carl Zeiss Vision International GmbH, Turnstraße 27, 73430 Aalen }

\maketitle
\noindent
MGG Orcid 0000-0001-7379-0080\\
YS Orcid 0000-0002-7513-341X\\
TW Orcid 0000-0001-7899-0858\\
SW Orcid 0000-0003-3437-6711\\
\small{Corresponding author: Miguel García García\\
miguel.garcia-garcia@uni-tuebingen.de}
\begin{abstract}
For decades, manufacturers have attempted to reduce or eliminate the optical aberrations that appear on the progressive addition lens' surfaces during manufacturing. Besides every effort made, some of these distortions are inevitable given how lenses are fabricated, where in fact, astigmatism appears on the surface and cannot be entirely removed or where non-uniform magnification becomes inherent to the power change across the lens. Some presbyopes may refer to certain discomfort when wearing these lenses for the first time, and a subset of them might never adapt. Developing, prototyping, testing and purveying those lenses into the market come at a cost, which is usually reflected in the retail price. This study aims to test the feasibility of virtual reality for testing customers' satisfaction with these lenses, even before getting them onto production. VR offers a controlled environment where different parameters affecting progressive lens comforts, such as distortions, image displacement or optical blurring, can be analysed separately. In this study, the focus was set on the distortions and image displacement, not taking blur into account. Behavioural changes (head and eye movements) were recorded using the built-in eye tracker. Participants were significantly more displeased in the presence of highly distorted lens simulations.
In addition, a gradient boosting regressor was fitted to the data, so predictors of discomfort could be unveiled, and ratings could be predicted without performing additional measurements.

\textbf{Keywords: virtual \and reality \and distortions \and comfort \and progressive \and addition \and lenses \and eye-tracking}
\end{abstract}

\setlength{\parskip}{0em}
\section{Introduction} \label{Introduction}
\acrfull{Pals} provide presbyopes with clear vision at different distances, with the use of a single ophthalmic lens \citep{Poullain1911OpticalLens.}. Firstly patented in 1907 \citep{Aves1907SpecialSurfaces}, multifocal ophthalmic lenses have undergone a great deal of evolution \citep{Sullivan1988ProgressiveReview}. Although their introduction to the market was neither rapid nor widely accepted, being only in 1959, when precursors of modern progressive lenses became commercially available \citep{Maitenaz1969OphthalmicPower, VOLK1962ThePresbyopia}, over the years, progressive lenses have become the most popular solution for presbyopia.

In general, the fabrication of progressive power lenses leaves inevitably residual surface astigmatism that leads to peripheral distortions \citep{Minkwitz1963UberAspharen, Esser2017GeneralizationSurfaces.}. How this residual astigmatism is spread across the surface was usually employed to classify \acrshort{pals} into soft (more spread) and hard designs (more concentrated) \citep{Atchison1987OpticalLenses}. However, nowadays, these lenses navigate between both classes, diffusing the borders as to whether they belong to one or another category, with manufacturers aiming to obtain the best combination of parameters for each group of individuals.

Besides their popularity and the fact that they do seem to improve wearers' quality of life \citep{AhmadNajmee2017SatisfactionWearers}, not everyone gets used to them. Scientific literature is scarce when it comes to gathering knowledge about why few "rookie" wearers suffer in adapting to these lenses or which could be the main factors contributing to this. Some papers mainly refer to spectacles in general and report errors in the prescription as the primary source of discomfort \citep{Bist2021}. However, assuming no errors from the optician, a small group still reports discomfort due to distortions and multifocality. \citet{Alvarez2017AdaptationConsider} proposed that non-adopters may have a weaker ability to modify their convergence and a reduced rate and magnitude of phoria adaptation. Yet, individuals who reject these lenses report blurred vision, headaches, perceived movement of the peripheral visual field (a.k.a. \textit{swim}), balance issues and even nausea \citep{ChoM1991, Han2003DynamicEnvironment}, symptoms more likely associated with poor adaptation to novel visual conditions \citep{Alvarez2009AdaptationPresbyopes}.

Traditionally, as reported by \citet{Ogle1950ResearchesVision}, one of the most significant handicaps when analysing the tolerance to distortions induced by ophthalmic lenses was the difficulty in separating them from blur \citep{Barbero2020AdmissibleLenses}. With the computational development of the last decades, it is entirely possible to present an image that is solely distorted \citep{Habtegiorgis2017AdaptationAftereffects, Sauer2020ParallelDistortions} or blurred \citep{Sawides2010AdaptationBlur, Vinas2012PerceptualAstigmatism}, helping to understand how our visual system adapts to these changes. However, these experiments are somewhat forced, built upon certain constraints, such as specific gazing or a fixed head position, and while they do drive knowledge forward, they often lose sight of the big picture.

On the other hand, one can analyse eye-tracking data while wearing \acrshort{pals} \citep{Hutchings2007EyeWearers}, bringing more natural conditions at the expense of not fully comprehending which features are affecting what since they can not be separated. In addition, the potential source of error that arises when fitting \acrshort{pals} cannot be ignored. All in all, virtual reality offers a perfect test bench as it can replicate the effect of distortions and blurring apiece while maintaining the freedom of movement and immersion of virtual environments.

This study aims to test the feasibility of virtual reality to assess, benchmark and rate the discomfort that four different ophthalmic lenses with various refractive powers can introduce, given the distortions they present. Subjective grading was performed continuously after finishing a simple task involving navigation, locomotion, scene exploration, and grasping to recognise the subjects' ability to perceive such simulated distortions and assess their comfort when affected by them. Possible behavioural changes prompted by optical distortions were also sought, and a model was developed to predict what the discomfort might be like without requiring additional measurements.
\section{Material \& Methods} \label{MaterialAndMethods}
\subsection{Subjects and informed consent}
A total of 18 naïve subjects (9 males / 9 females) participated in the course of the study. The participants were aged between 19 and 30 years (mean = 24; SD = 3). None of the subjects presented a prior history of problems using a virtual reality headset or motion issues, nor did they present any refractive error that could have compromised their vision during the experiment or influenced their perception due to the usual wear of their lenses.

\subsection{Ethics}
The study adhered to the tenets of the Helsinki Declaration (2013) and subsequent amends. The ethics authorisation to perform the measurements was granted by the Ethics Committee at the Medical Faculty of the Eberhard-Karls University and the University Hospital Tübingen with the ID 986/2020BO2. Before data collection, the experiment was explained in detail to the participants, and written informed consent from each participant was stored. All data was pseudo-anonymised and stored in full compliance with the principles of the Data Protection Act GDPR 2016/679 of the European Union \citep{TheEuropeanParliamentandtheCounciloftheEuropeanUnion2016RegulationCouncil}.

\subsection{Set-Up}
\begin{figure}[h!]
\begin{center}
\includegraphics[width=0.5\textwidth]{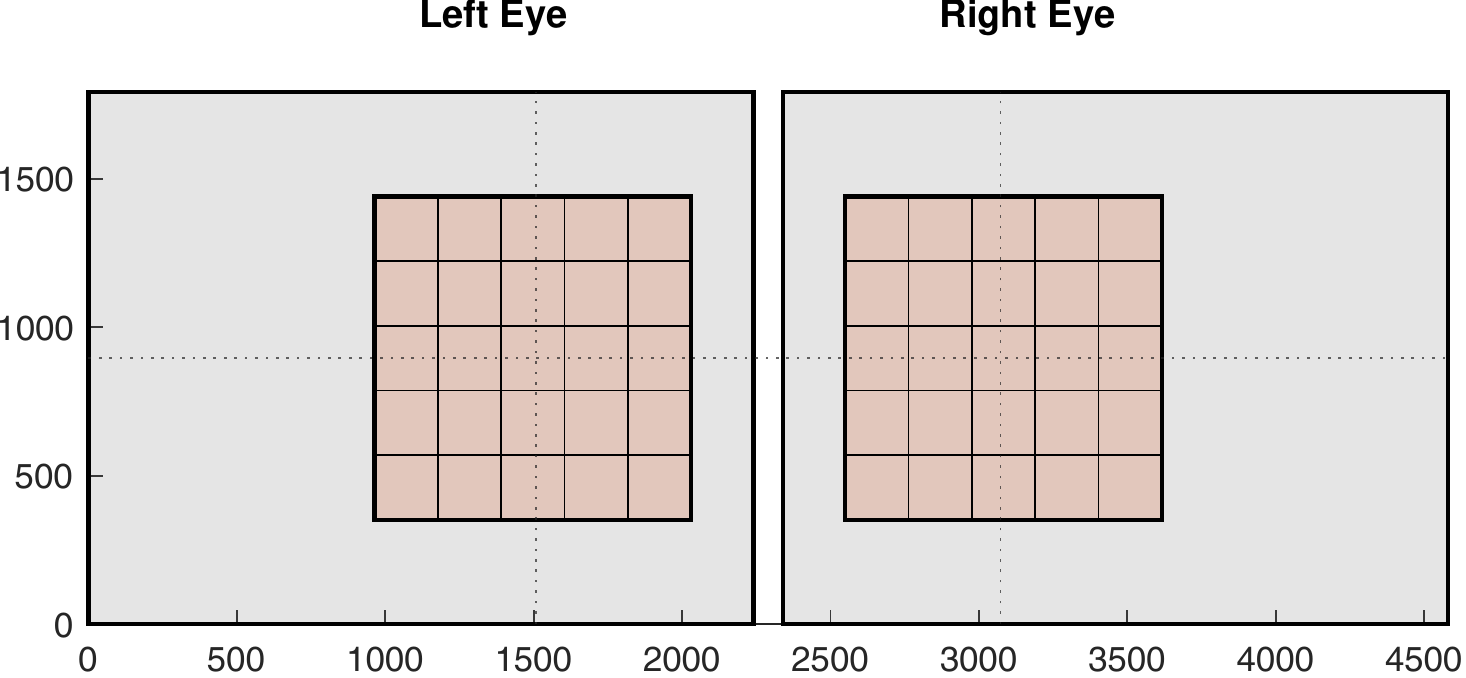}
\end{center}
\caption{Area of display resolution of the VR headset(gray) as well as area in pixels covered by the eye tracker(orange) as provided by the manufacturer.} \label{fig:1}
\end{figure}

A cuboidal booth of $\SI[]{3}{} \times \SI[]{3}{\meter}$ defined the physical space around which the participants were entitled to freely move. On every upper corner (4 in total), one HTC Vive lighthouse base station tracker v2 (HTC Cooperation, Xindian, Taiwan) recorded the position and orientation of the headset as well as the controller used. The StarVR One (StarVR, Taipei, Taiwan) head-mounted display was used to present the virtual environment to the participants. This \acrfull{hmd} has an effective \acrfull{fov} of \SI{182}{\degree} by \SI{99}{\degree} \citep{Sauer2022AssessmentTesting}, and an average eye relief (or distance from lens/display to the eye of \SI{18.2}{\mm} (15.8 to \SI{23.9}{\mm}). The display has a resolution of $\SI{2240}{\px} \times \SI{1792}{\px}$, a maximum luminance of \SI{68.4}{\candela\per\square\meter} and the refresh rate fixed at \SI{90}{\Hz}. This device presents eye-tracking capabilities through Tobii Pro libraries. For reference, the display area covered by the eye-tracker, as reported by the manufacturer, is plotted in Figure \ref{fig:1}. To provide input and locate the hand position, a standard HTC Vive controller was used. 

\subsubsection{Virtual environment \& Computer}
\begin{figure*}[h!]
\begin{center}
\includegraphics[width=\textwidth]{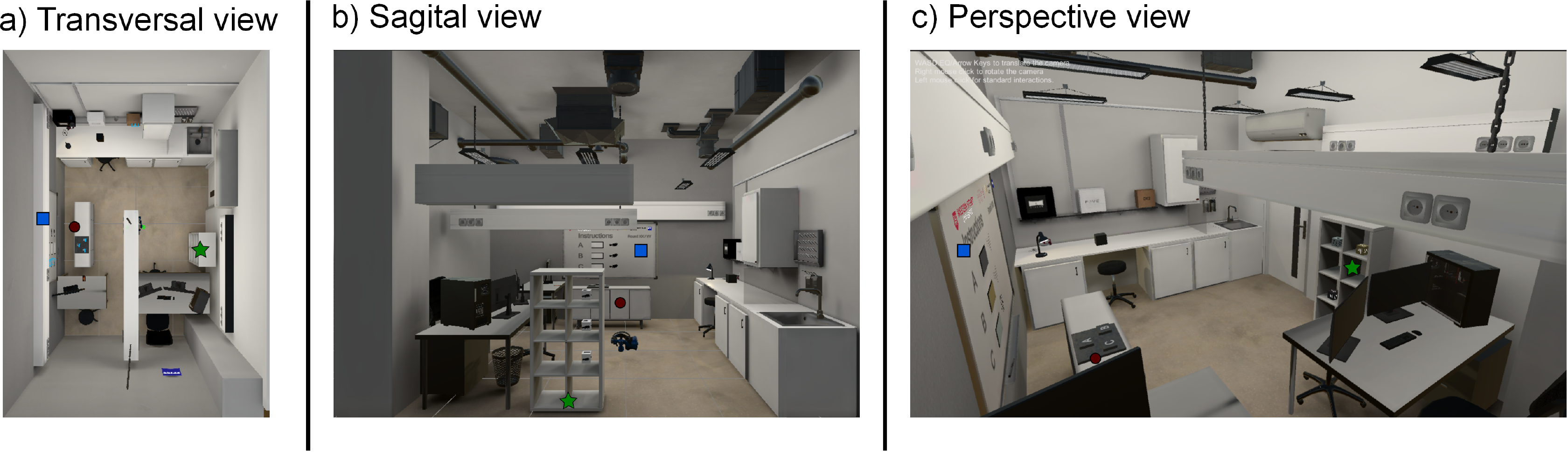}
\end{center}
\caption{View of the virtual environment. The blue square denotes where the instructions were located in the virtual environment, and the rating \acrfull{UI} was presented. The red circle is located where the cubes need to be placed. Finally, the green star is placed upon the cube spawns' origin.} \label{fig:2}
\end{figure*}

The virtual environment was constructed using the rendering engine Unity version 2019.4.25.f1 (Unity Technologies, California, USA) and Blender 3.0 (Stichting Blender Foundation, Amsterdam, Netherlands). The experiment was written in C\# and the eye-tracking data was recorded using the Tobii Pro (Tobii Pro AB, Danderyd, Sweden) eye-tracking libraries. A view of the virtual reality environment used is depicted in Fig \ref{fig:2}.

The whole experiment ran on a Windows 10(20H2) PC with an Intel(R) Core i7-10700, 16 GB of RAM and an Nvidia 3080 GPU.

\subsection{The task}
After providing informed consent and having the experiment explained in detail, participants had time to ask questions. They then wore the virtual reality headset to acclimate to the scene by wandering around the virtual environment (a replica of a lab room) before the actual assignment started. Across the task, participants were requested to position three different virtual cubes from point A (Green star) to point B (Red circle)(See Fig \ref{fig:2}). The location at which those cubes spawned was randomised between several pre-defined positions in the shelf unit. The cubes' colours, the patterns they presented on their surface and the location where they were supposed to be placed were also randomly changed across trials. Instructions on where the cubes had to be placed were provided on a virtual whiteboard (Blue square) opposite the cubes' spawn origin. Thus, participants were forced to move and look around in the virtual environment during each trial.

\subsubsection{Ratings of discomfort}
After every cube was correctly placed on the plate, a \acrshort{UI} panel with a slider popped up in front of the participant. In this panel, the participants were asked to report their level of visual discomfort on a scale from 0 (no discomfort) to 7 (highest discomfort). Previous to providing this feedback, participants were instructed to rate only based on how the distortions affected their perception of their visual comfort and to not take into account factors such as possible frame drops or the image resolution. Given the rating and confirmation, the trial was completed, and a new trial started.

A total of 25 trials were answered (five per lens condition), with an average duration of a trial of \SI{36}{\s} (SD = \SI{12}{\s}, range from \SI{18}{\s} to \SI{101}{\s}).

\subsection{Lens Conditions - Distortion simulation}
Four progressive additional lenses of the same type, but different prescription power were used. Different addition powers or spherical power prescriptions do modify how distortions are present. The distortions profiles were provided by the manufacturer from a set of lenses with the following parameters fixed: a refractive index of 1.5, a pantoscopic angle of \SI{7.5}{\degree}, a \SI{5.28}{\mm} base curvature, a back vertex distance of \SI{12}{\mm} and a corridor length of \SI{14}{\mm}. Two of these lenses had non spherical power and addition powers from +2 and \SI[retain-explicit-plus]{+3}{\diopter}, and the other two lenses had an addition power \SI[retain-explicit-plus]{+2}{\diopter} and a spherical refractive correction of +2 and \SI{-2}{\diopter}. In line with modern \acrshort{pals}, these lenses have their "umbilical line" tilted, with the near point shifted towards the nasal side to account for vergence. 

A set of grid points ($x$, $y$) on a plane perpendicular to a straight gaze direction is perceived in a distorted position ($x_{d}$, $y_{d}$) when seen through a progressive addition lens. The displacement between ($x_{d}$, $y_{d}$) and ($x$, $y$) was pre-computed for each lens using ray tracing based on these lenses' digital lens surface data. The displacement in the \textit{horizontal} and \textit{vertical} coordinates were encoded into a raster image (EXR format) where the red colour channel was used for the shift of the pixel in the horizontal direction and the green colour for the vertical component.

\begin{figure*}[htpb]
\begin{center}
\includegraphics[width=\textwidth]{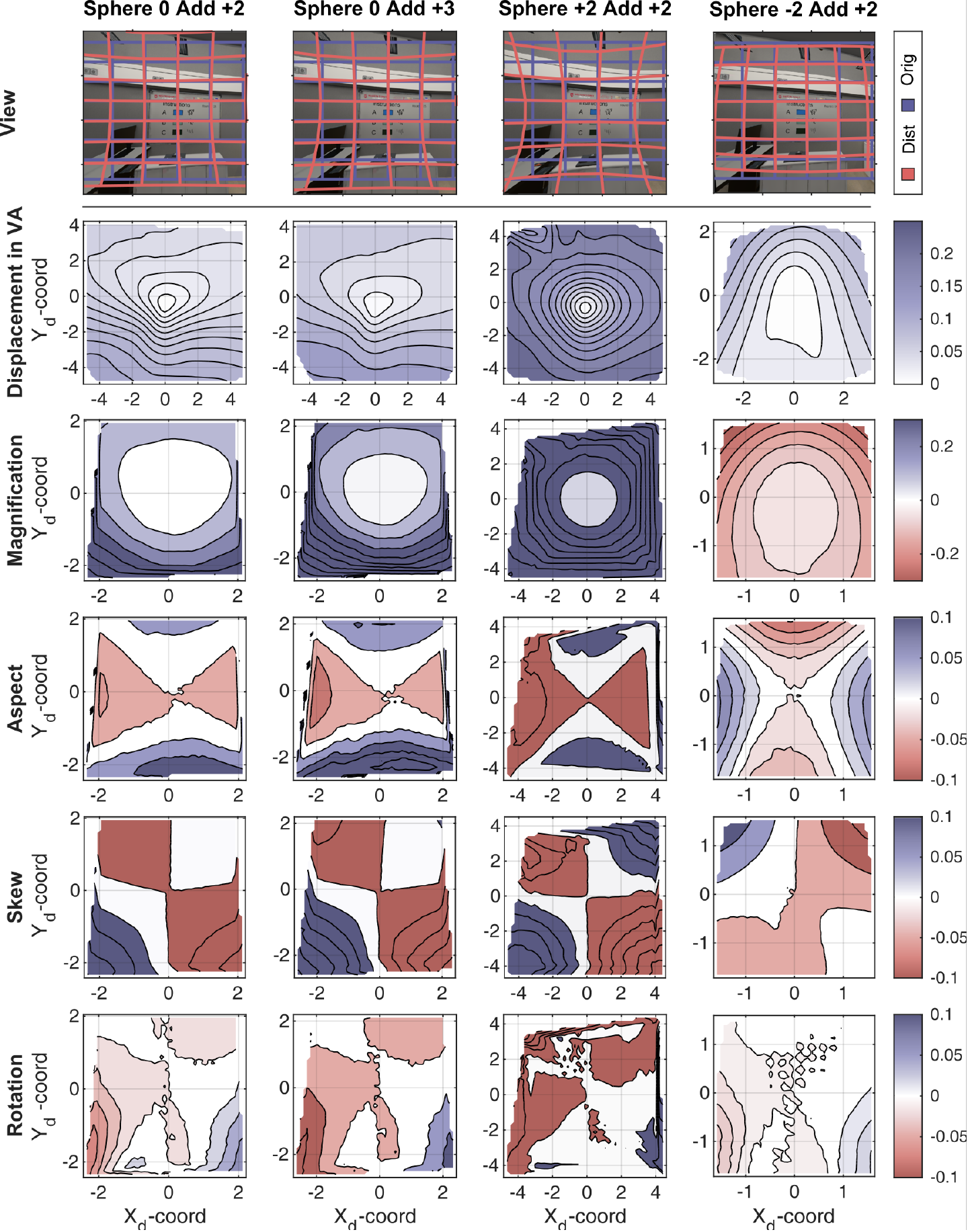}
\end{center}
\caption{Representation of how distortions affect a grid and the image observed (background), and maps of pixel displacement in visual angles as observed from the subject's perspective for the OD, as well as magnification, differences in aspect ratio, skew and rotation calculated from the first derivatives of the distortions.} \label{fig:3}
\end{figure*}

These distortions were then applied pixel-wise to the image displayed in the virtual reality headset through a custom shader in Unity 3D, written in \textit{HLSL}. The amount of pixel displacement that these lens simulations prompted in visual angle, along with the magnification, skew, aspect or rotation, as seen by the participants, can be observed in Fig \ref{fig:3}.

The angular displacement in the visual field as observed was computed using the following formula:
\begin{align}
\begin{split}
   \mathrm{Displacement \hspace{.1cm} (V_{(Visual \hspace{.1cm} angle \hspace{.1cm} in \hspace{.1cm} degrees )})} =\\
   cos^{-1} \left(\frac{x_d \times x + y_d \times y +1}{\sqrt{(x_d^2 + y_d^2+1)\times(x^2 +y^2 +1)}} \right);
\end{split}
\end{align}
Where ($x$, $y$) are the coordinates of the original grid points and ($x_{d}$, $y_{d}$) are the location of those points after the distortion is applied.

\subsection{Tracking}

On every frame, the position coordinates of the \acrfull{hmd} along with the rotation quaternions and gaze vectors were stored.
\subsubsection{Head-tracking}
Head rotation values were converted using \emph{quat2eul} into Euler angles following the MATLAB coordinates system for analysis and plotting. The table \ref{tab:Coordinates systems} summarises the different coordinates systems that are used in the main libraries of this study. 

\begin{table}[H]
\SetTblrInner{vspan=minimal}
\begin{tblr}[]{Q[c,m]|Q[c,m]|Q[c,m]Q[c,m]Q[c,m]|Q[c,m]}
\toprule
{Program \\Library} & {Coordinate \\ System} & X & Y & Z & {Euler angles \\order} \\
\midrule
 Unity 3D  & Left-handed & $\rightarrow$ & $\uparrow$ & $\nearrow$ & z,x,y \\
 MATLAB & Right-handed & $\nearrow$ & $\leftarrow$ & $\uparrow$  & z,y,x\\
\bottomrule
\end{tblr}
\caption{Coordinates system references for each software system. $\rightarrow$ - right; $\leftarrow$ - left; $\uparrow$ - up; $\nearrow$ - forward.}
\label{tab:Coordinates systems},
\end{table}

On every trial the \textit{yaw}, \textit{pitch} and \textit{roll} were computed using the \acrshort{hmd} quaternions as described in the equation \ref{eq:01}.
\begin{align}
\begin{split}
\mathrm{Tait \hspace{.1cm}Bryan \hspace{.1cm} Angles\hspace{.1cm} (roll, pitch, yaw) \hspace{.1cm}} = \\
\textit{quat2eul}(rotW, rotZ, -rotX, rotY, XYZ);
\end{split}
\label{eq:01}
\end{align}

The roll data (tilt of the head) was further characterised with a \textit{Von Misses} distribution fit (See equation \ref{eq:vonmis}). 

\begin{align}
\begin{split}
    \mathcal{M}\hspace{.1cm}(\theta | \hspace{.1cm}\mu,\kappa ) = {\left[ 2\pi I_{0}(\kappa)\right]}^{-1} exp\{\kappa \hspace{.1cm} cos(\theta-\mu)\}; \\ \label{eq:vonmis}
    \end{split} 
\end{align}
Where $\mu$ is the mean direction, $I_{0}$ is the \textit{bessel} function of the first kind, $\kappa$ is the concentration parameter, and $\theta$ refers to the angle in radians from $-\pi$ to $\pi$.
 
The pitch (or head inclination up to down) was described using a combination of two \textit{Von Misses} distributions. And finally, the yaw (or horizontal head rotation) was represented by the fit (See equation \ref{eq:pb}) of two \textit{inverse Power Batschelet} distributions \citep{Mulder2019BayesianProblems,Mulder2020MixturesDirections}, as the data was highly peaked towards the cube spawns origin and the horizontal location of the plate, whiteboard and rating \acrshort{UI} panel.
 
\begin{align}
\begin{split}
    f_{\mathcal{PB}}\hspace{.1cm}(\theta | \hspace{.1cm}\mu,\kappa, \lambda ) = \left[ \mathcal{K}_{\kappa, \lambda}^{\ast}\right]^{-1} exp\{\kappa \hspace{.1cm} cos \hspace{.1cm} t_{\lambda}^{\ast} (\theta-\mu)\};\\
    \mathcal{K}_{\kappa, \lambda}^{\ast} = \int_{-\pi}^{\pi} exp\{\kappa \hspace{.1cm} cos \hspace{.1cm} t_{\lambda}^{\ast} (\theta-\mu)\} \hspace{.1cm}d\theta ;\\
    t_{\lambda}^{\ast} \hspace{.1cm}(\theta) = sign(\theta) \pi \left( \frac{|\theta|}{\pi}\right)^{\gamma(\lambda)};\\
    \gamma(\lambda) = \frac{1- c \lambda}{1+ c \lambda};\\ \label{eq:pb}
    \end{split} 
\end{align}
where c = 0.04082284 as noted by \citet{Mulder2019BayesianProblems}.

\subsubsection{Eye Tracking}
The eye-tracking data was obtained using the Tobii Pro SDK 1.10 for Unity. The SDK was modified to further record the hit-point of the gaze vector in the world coordinates. 

\paragraph*{Fixations}
While recording the eye-tracking data, on every frame, the combined (head and gaze) origin and direction were used to cast a ray within Unity until it hit a collider in the scene. The coordinates of this hit-point (x,y,z) were later used to detect fixations. If a cluster of continuous hit-points stood within the volume of a sphere of \SI[]{0.05}{\meter} radius, during more than \SI[]{200}{\ms} (See \citet{VanDerLans2011} for a review on average fixations duration while visual search and scene viewing), it was considered a fixation. The amounts of fixations performed per minute and their average duration were recorded on every trial.

\paragraph*{Saccades}
To estimate saccadic eye movements, the gaze ray from the right eye (in headset relative coordinates) was transformed into polar ($\theta$) and azimuthal angles ($\phi$). The median angular distance travelled ($\psi$) between the original frame vector, and the five subsequent frames was computed following the equation \ref{eq:psi}. Then, the angular speed was estimated as the angular distance per second, on which a Savitzky-Golay filter \citep{Dai2016AMovement} 
was applied.
\begin{align}
    \begin{split}
    \phi = \arctan \left( \frac{y}{x} \right); \\
    \theta = \arccos (z); \\
    \psi = \\\arccos \Bigl( (\sin(\theta_{n+i}) \times \sin(\theta_{n}) \times \cos(\phi_{n+i}-\phi_{n}) \\
    + \cos(\theta_{n}) \times \cos(\theta_{n+i}) \Bigr); \\
    \end{split}
    \label{eq:psi}
\end{align}

A speed threshold of \SI[]{50}{\degree \per \second} and a minimum distance ($\psi$) of \SI[]{1}{\degree} were defined as thresholds for defining a saccade. The number of saccades per minute, the average amplitude and maximum peak velocity performed on every trial were recorded.

\paragraph*{Eye tracking to measure average observed displacement, magnification, aspect, skew and rotation}
Every gaze position was projected over the displacement, magnification, skew, rotation, and differences on aspect ratio maps, and the observed absolute and relative ("foveally") pixel displacement, magnification, skew, rotation and aspect ratio parameters were computed by summing the values at gaze position over all the frames of the trial. The mean value and standard deviation per trial were recorded and compared.

\section{Analysis \& Preprocessing of the Data} \label{Analyisis}
All the statistical analysis and plotting were performed in MATLAB 2020b (Mathworks Inc, Natick, CA, USA) and Python 3.10 with sci-kit learn 1.0.2 \citep{Pedregosa2011Scikit-learn:Python}.

\section{Results} \label{Results}
\subsection{Visual discomfort}
On every trial, the participants rated their visual discomfort on a scale from 0 to 7, with zero being normal and seven being the most uncomfortable. Discomfort is, in fact, a subjective parameter that is highly susceptible to vary between subjects and depends on where the participant sets the threshold towards unpleasantness. Given the same distortion and conditions, two subjects can be uncomfortable or not. To standardise our answers and limit the subject's susceptibility, the ratings were normalised using the ratings where no lens distortion was presented as a baseline. The formula used can be observed in the equation \ref{eq:02}.
\begin{align}
\begin{split}
R_{\{i\}} \hspace{.1cm} = R_{\{i\}} \hspace{.1cm} orig \times \frac{10}{7} + 1; \\
R_{\{i\}} \hspace{.1cm} norm = \frac{R_{\{i\}}}{\overline{R}_{\{baseline\}}}; 
\end{split}
 \label{eq:02}
\end{align}

\begin{figure}[]
\begin{center}
\includegraphics[width=\columnwidth]{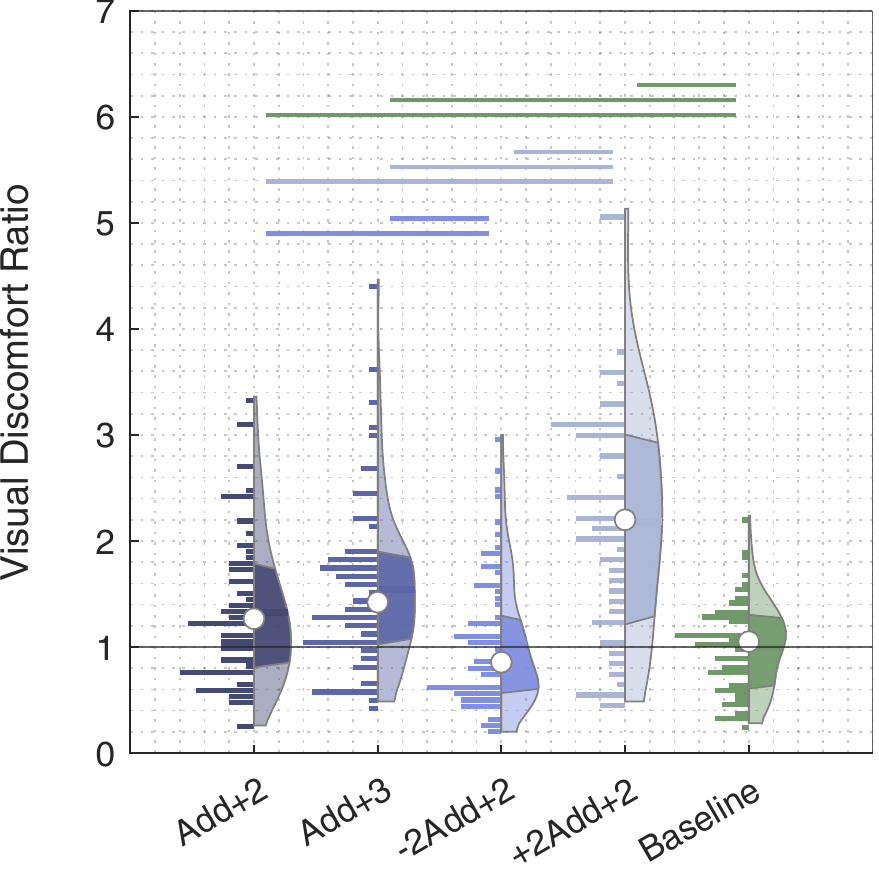}
\end{center}
\caption{Visual discomfort ratings' distribution across conditions, all values have been normalised. The horizontal bars on top indicate significant differences in Post-Hoc analysis.}
\label{fig:Exp02_Visual_Distortions_Distribution}
\end{figure}

The distribution of visual discomfort ratings was not parametric in any of the conditions tested. Thus, a non-parametric test (Kruskal-Wallis) was used to compare lens condition discomfort ratings. Significant differences were found between the distributions of visual discomfort ratings across conditions (p $<$ 0.01; $\tilde{\chi}^2$ = 69). Bonferroni's Post-Hoc test indicated that the lens distortions from the lens +2D and addition +2 were more disturbing than the rest. Likewise, the lens with a spherical power of -2 and addition +2 had the smallest discomfort. Fig \ref{fig:Exp02_Visual_Distortions_Distribution} shows the distribution of ratings and the statistical differences.

The Cronbach`s alpha value ($\rho_{\tau}$) for the rating scale was found to be statistically consistent (0.88).
\subsection{Gender \& Age}
No bias was found in the ratings of discomfort due to gender (Mann-Withney-U; $N_{females}$ = 9 (MdN = 0.29); $N_{males}$ = 9 (MdN = 0.29); U = 51917 ; z-val = 0.87; p = 0.39). 

No correlation was found between the ratings and the age of the participants (Spearman - $R^{2}$ = 0.0018; p = 0.97).

\subsection{Duration}
A small correlation ($R^{2}$ = 0.11, p = 0.02) was found between the duration of the trials and the rating given. In general, subjects seemingly were faster deciding whether to give the highest and the lowest ratings of discomfort. However, subjects may have taken longer times during the trial if the decision was unclear, as they perceived some level of discomfort but not the highest.

\subsection{Yaw, pitch and roll}
Tables \ref{tab:pitchtable}, \ref{tab:yawtable} and \ref{tab:rolltable} present the main characteristics of the fitted distributions for pitch, yaw and roll per trial with their mean and standard deviations. The statistical differences of each parameter (individual, lens or rating (rounded to the closest integer)) were computed using the Kruskal-Wallis test, and are also reported in the tables. 

\begin{table*}[p]
    \centering
    \includegraphics[width= \textwidth]{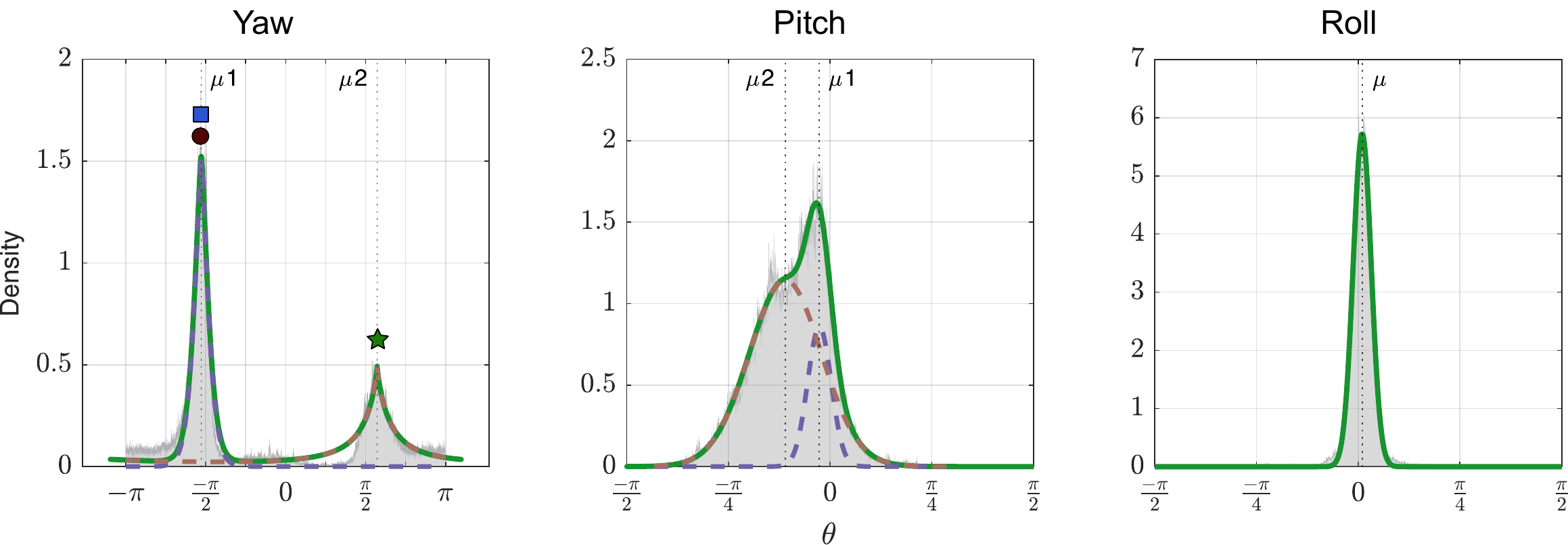}
    \captionof{figure}{Examples of distributions fitted for Yaw, Pitch and Roll. The green line denotes the final distribution fitted, and the dashed red and blue lines indicate the individual distributions (inverse power Batschelet or von Mises, contributing to the final one). The icons on the Yaw plot represent the same locations as in the Fig. \ref{fig:2})}
    \label{fig:yprfigs}
    
    \begin{tblr}{width= \textwidth,colspec={Q[c,m]|X[c,m]X[c,m]X[c,m]X[c,m]X[c,m]|Q[c,m]Q[c,m]Q[c,m]}}
    \toprule
    &$-/+2$ & $-/+3$ & $-2/+2$ & $+2/+2$ & Baseline & Lens & Ratings & Subj\\
    \midrule
    $\mu_{1}$ & $-1.65 \pm 0.10$ & $-1.64 \pm 0.11$ & $-1.65 \pm 0.12$ & $-1.65 \pm 0.10$ & $-1.66 \pm 0.08$ & $n.s.$ & $n.s.$ &$***$\\
    $\kappa_{1}$ & $\hspace{1ex}6.04 \pm 2.46 \hspace{1ex}$&$\hspace{1ex} 6.58 \pm 2.88 \hspace{1ex}$& $\hspace{1ex} 6.42 \pm 2.68 \hspace{1ex}$&$ \hspace{1ex}6.57 \pm 2.91 \hspace{1ex}$&$\hspace{1ex} 6.15 \pm 2.54\hspace{1ex} $& $n.s.$ & $n.s.$ &$***$\\
    $\lambda_{1}$ & $0.80 \pm 0.22$ & $0.76 \pm 0.27$ & $0.75 \pm 0.26$ & $0.68 \pm 0.31$ & $0.77 \pm 0.23$ & $n.s.$ & $n.s.$ &$***$\\
    FWHM$_{1}$ & $0.20 \pm 0.12$ & $0.23 \pm 0.17$ & $0.23 \pm 0.17$ & $0.28 \pm 0.20$ & $0.22 \pm 0.12$ &$**$&$***$&$***$\\
    $\omega$ & $0.59 \pm 0.09$ & $0.58 \pm 0.11$ & $0.59 \pm 0.10$ & $0.60 \pm 0.10$ & $0.59 \pm 0.09$ & $n.s.$ & $n.s.$ &$***$\\
    $\mu_{2}$ & $1.75 \pm 0.13$ & $1.77 \pm 0.13$ & $1.75 \pm 0.14$ & $1.77 \pm 0.17$ & $1.76 \pm 0.16$ & $n.s.$ & $n.s.$ &$***$\\
    $\kappa_{2}$ & $2.10 \pm 0.76$ & $2.17 \pm 0.99$ & $2.26 \pm 0.96$ & $2.25 \pm 0.84$ & $2.10 \pm 0.72$ & $n.s.$ & $n.s.$ &$***$\\
    $\lambda_{2}$ & $1.00$ & $1.00$ & $1.00$ & $1.00$ & $1.00$ & $n.s.$ & $n.s.$ &$***$\\
    FWHM$_{2}$ & $0.34 \pm 0.18$ & $0.35 \pm 0.21$ & $0.32 \pm 0.18$ & $0.32 \pm 0.23$ & $0.34 \pm 0.18$ & $n.s.$ & $n.s.$ &$***$\\
    \bottomrule
    \end{tblr}
    \captionof{table}{Yaw. Two inverse power Batschelet $\mathcal{F}$ PB were fitted. In addition to those in the caption of Table 2, $\lambda$ refers to the peakness of the fitted distribution.\\}
    \label{tab:yawtable}
    
    \begin{tblr}{width= \textwidth,colspec={Q[c,m]|X[c,m]X[c,m]X[c,m]X[c,m]X[c,m]|Q[c,m]Q[c,m]Q[c,m]}}
    \toprule{1-9}
    &$-/+2$ & $-/+3$ & $-2/+2$ & $+2/+2$ & Baseline & Lens & Ratings & Subj\\
    \midrule
    $\mu_{1}$ &  $-0.05 \pm 0.39$ & $-0.06 \pm 0.39$ & $-0.03 \pm 0.48$ & $-0.10 \pm 0.42$ & $-0.01 \pm 0.53$ & $n.s.$ & $*$ &$***$ \\
    $\kappa_{1}$ & $277.66 \pm 145.80$ & $286.41 \pm 160.89$ & $283.69 \pm 148.15$ & $296.62 \pm 150.39$ & $327.90 \pm 155.98$ & $n.s.$ & $*$ &$***$ \\
    FWHM$_{1}$ &  $0.55 \pm 0.19$ & $0.55 \pm 0.22$ & $0.55 \pm 0.21$ & $0.57 \pm 0.21$ & $0.54 \pm 0.20$ & $n.s.$ & $n.s.$ &$***$ \\
    $\omega$ & $0.19 \pm 0.22$ & $0.20 \pm 0.20$ & $0.15 \pm 0.18$ & $0.19 \pm 0.20$ & $0.17 \pm 0.19$ & $n.s.$ & $n.s.$ &$***$ \\
    $\mu_{2}$ & $-0.33 \pm 0.15$ & $-0.33 \pm 0.17$ & $-0.31 \pm 0.14$ & $-0.32 \pm 0.14$ & $-0.33 \pm 0.16$ & $n.s.$ & $*$ &$***$ \\
    $\kappa_{2}$ & $48.66 \pm 90.70$ & $49.47 \pm 83.45$ & $45.12 \pm 88.38$ & $44.28 \pm 83.99$ & $48.41 \pm 79.49$ & $n.s.$ & $n.s.$ &$***$ \\
    FWHM$_{2}$ & $0.52 \pm 0.31$ & $0.45 \pm 0.29$ & $0.57 \pm 0.32$ & $0.49 \pm 0.29$ & $0.50 \pm 0.36$ & $n.s.$ & $n.s.$ &$***$ \\
    \bottomrule
    \end{tblr}
    \caption{Pitch. Two Von Mises ($\mathcal{M}$) were fitted. For each of them $\mu$ stands for mean direction, $\kappa$ is concentration, $\omega$ is the contribution of the function from 0 to 1, and \textit{FWHM} stands for full-width half maximum of the distribution. 
    \\}
    \label{tab:pitchtable}

    \begin{tblr}{width= \textwidth,colspec={Q[c,m]|X[c,m]X[c,m]X[c,m]X[c,m]X[c,m]|Q[c,m]Q[c,m]Q[c,m]}}
    \toprule
    &$-/+2$ & $-/+3$ & $-2/+2$ & $+2/+2$ & Baseline & Lens & Ratings & Subj\\
    \midrule
    $\mu$ &  $0.02 \pm 0.04$ & $0.02 \pm 0.05$ & $0.03 \pm 0.04$ & $0.02 \pm 0.04$ & $0.03 \pm 0.04$ & $n.s.$ & $n.s.$ &$***$\\
    $\kappa$ & $344.33 \pm 142.43$ & $352.14 \pm 128.91$ & $360.01 \pm 130.41$ & $361.83 \pm 125.27$ & $382.43 \pm 127.06$ & $n.s.$ & $**$ &$***$\\
    FWHM & $0.14 \pm 0.04$ & $0.14 \pm 0.04$ & $0.13 \pm 0.04$ & $0.13 \pm 0.03$ & $0.13 \pm 0.03$ & $n.s.$ & $**$ &$***$\\
    \bottomrule
    \end{tblr}
    \caption{Roll. A single Von Mises distribution ($\mathcal{M}$). \\}
    \label{tab:rolltable}
\end{table*}

None of the parameters above mentioned was significantly different across lens conditions, but the full-width half maximum (FWHM) of the first inverse power Bachelet distribution on yaw. The same FWHM, as well as the mean directions of pitch, and the FWHM of the roll changed across rating groups. 

\subsection{Gaze behaviour}

\paragraph*{Fixations}
The number of fixations per minute and duration of fixations were significantly different across subjects (Kruskal-Wallis, H(17) = 255, p $<$ 0.001) and (H(17) = 197, p $<$ 0.001). Significant differences were also found in the number of fixations per minute across ratings (KW, H(5) = 25.7, p$<$0.001), where the greater discomfort is perceived, the higher amount of fixations are found, with the exception of the group with the highest discomfort (4-5 ratings).

\paragraph*{Saccades}
The number of saccades per minute, as well as the peak velocity and the average distance, travelled on a saccade, were significantly different across participants (Kruskal-Wallis, H(17) = 269, p $<$ 0.001),(KW, H(17) = 301, p $<$ 0.001) and (KW, H(17) = 285, p $<$ 0.001). In the trials with more discomfort, the saccades were on average larger (KW, H(5) = 14, p $<$ 0.05) and had a higher peak velocity (KW, H(5) = 18, p $<$ 0.01). Significantly (KW, H(4) = 9.6, p $<$ 0.05) more saccades were performed when the lens presented had higher amount of distortions. 

\paragraph*{Observed displacement, magnification, rotation, skew and aspect}
All the components that defined the experienced distortion computed from the movement data of each participant and the lens worn were significantly different (KW, H(4) < 434 , p $<$ 0.001) for every lens condition tested in their mean and standard deviation. The ranking order for it can be observed in Table \ref{tab:textureparamtable}.
Significantly more mean pixel displacement, magnification, differences on aspect ratio, skew, and rotation were observed when the subjects indicated higher discomfort (KW, H(5) < 103 , p $<$ 0.001). The standard deviation of all the observed parameters was significantly higher when more discomfort was perceived.
No differences across subjects were found for any of the mean values or standard deviation, but rotation standard deviation differences across subjects was close to significant (KW, H(17) = 27, p = 0.057).  

\begin{table}[H]
    \centering
    \definecolor{indigo}{rgb}{0.0, 0.25, 0.42}
    \begin{tblr}{width=\columnwidth, colspec={Q[l,m]|Q[c,m]Q[c,m]Q[c,m]Q[c,m]}}
    \toprule
    Parameter & $^{+2}/_{+2}$ & $^{-}/_{+3}$  & $^{-}/_{+2}$  & $^{-2}/_{+2}$ \\
    \midrule
    Displacement Mean  &  \SetCell{bg=indigo!85} &  \SetCell{bg=indigo!75} &  \SetCell{bg=indigo!50} &  \SetCell{bg=indigo!25} \\
    Displacement SD    &  \SetCell{bg=indigo!85} &  \SetCell{bg=indigo!50} &  \SetCell{bg=indigo!25} &  \SetCell{bg=indigo!75} \\
    Magnification Mean &  \SetCell{bg=indigo!85} &  \SetCell{bg=indigo!75} &  \SetCell{bg=indigo!50} &  \SetCell{bg=indigo!25} \\
    Magnification SD   &  \SetCell{bg=indigo!85} &  \SetCell{bg=indigo!75} &  \SetCell{bg=indigo!50} &  \SetCell{bg=indigo!50} \\
    Aspect Mean        &  \SetCell{bg=indigo!85} &  \SetCell{bg=indigo!75} &  \SetCell{bg=indigo!50} &  \SetCell{bg=indigo!25} \\
    Aspect SD          &  \SetCell{bg=indigo!85} &  \SetCell{bg=indigo!75} &  \SetCell{bg=indigo!50} &  \SetCell{bg=indigo!50} \\
    Skew Mean          &  \SetCell{bg=indigo!85} &  \SetCell{bg=indigo!75} &  \SetCell{bg=indigo!50} &  \SetCell{bg=indigo!25} \\
    Skew SD            &  \SetCell{bg=indigo!75} &  \SetCell{bg=indigo!50} &  \SetCell{bg=indigo!85} &  \SetCell{bg=indigo!85} \\
    Rotation Mean      &  \SetCell{bg=indigo!85} &  \SetCell{bg=indigo!50} &  \SetCell{bg=indigo!75} &  \SetCell{bg=indigo!75} \\
    Rotation SD        &  \SetCell{bg=indigo!75} &  \SetCell{bg=indigo!85} &  \SetCell{bg=indigo!25} &  \SetCell{bg=indigo!50} \\
    \bottomrule
    \end{tblr}
    \caption{Matrix draw showing the lens conditions ordered from highest (darker) to lowest(brighter), if the same colour appeared no significant differences were found between these two conditions.\\}
    \label{tab:textureparamtable}
\end{table}

\section{Model - Gradient Boosting Regressor} \label{Model}
Understanding what features might have influenced the decision for specific discomfort ratings can be a herculean task if deferred to traditional methods. However, thanks to the advances in computational power and machine learning algorithms, it is possible nowadays. Using \textit{scikit learn} package in Python \citep{Pedregosa2011Scikit-learn:Python}, we built a \acrlong{gbr} \citep{Friedman2002StochasticBoosting}, which ensembles several decision trees, using the residuals to fit the next iteration and weighting them to obtain a final model, capable of not only predicting new ratings given the estimators but also providing the relevance of the estimators.

The dataset used comprised a total of 450 trials, and contained age, gender, duration, baseline, maximum and minimum ratings of the subject, absolute mean and standard deviations of the "foveally" observed factors as described before, each of the parameters fitted to pitch, yaw and roll, and the already analysed data from the eye-tracking (i.e. fixations and saccades).

70\% of the dataset was used to train the model, and the remaining 30\% was only used to test the model's accuracy, i.e. the data was never shown to the model until the testing phase. The split was performed with stratification on subjects and lens conditions and testing for its robustness. The best hyper-parameters were decided using a bayesian optimisation search from \textit{skopt} \citep{Head2020Scikit-optimize/scikit-optimize}, and a repeated stratified group cross-validation fold within a pipeline.

The best features were selected from the model using the meta-transformer \textit{SelectFromModel} and their relevance to the model was estimated using the \textit{"SHAP"} or \textit{SHapley Additive exPlanation} values \citep{Lundberg2017APredictions}.
\begin{figure}[h]
    \begin{center}
    \includegraphics[width=\columnwidth]{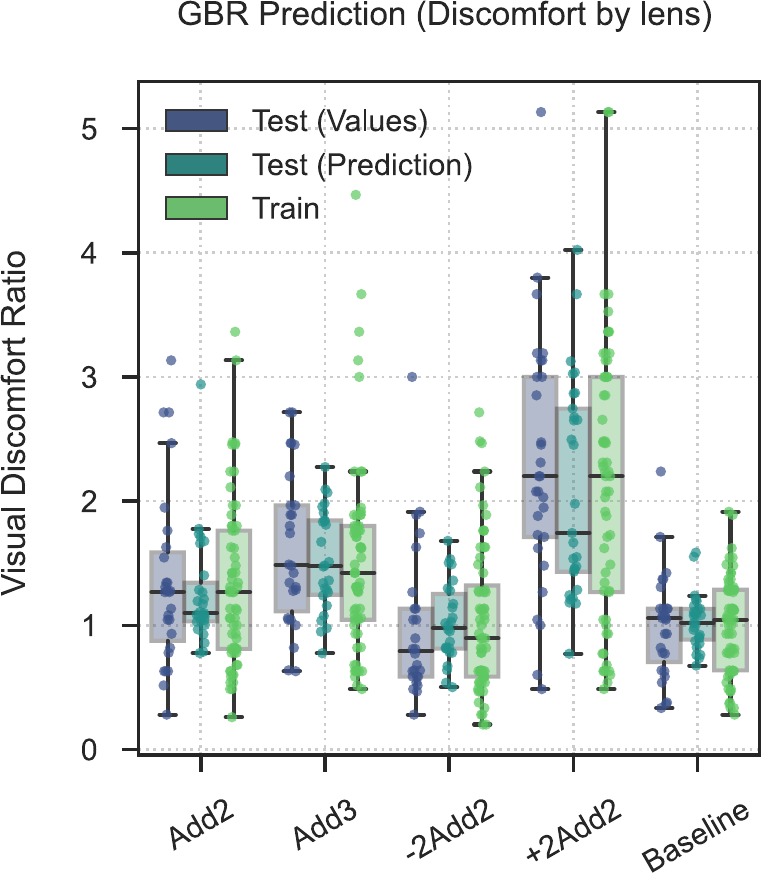}
    \end{center}
    \caption{Visual discomfort ratings' distribution across lenses, all values have been normalised. Different hues show the training dataset(true values), test dataset and the predictions from the model.}
    \label{fig:Predictions_Model}
\end{figure}

Over 400 seeds of random splits were tested to assure the model robustness, the table \ref{tab:Modelmetrics} shows the results of it.

\begin{table}[H]
\centering
 \begin{tblr}[]{width=\columnwidth, colspec= {Q[c,m]|Q[c,m]Q[c,m]Q[c,m]Q[c,m]Q[c,m]}}
    \toprule
     & $\mathcal{R}^2_{test}$ & $\mathcal{R}^2_{train}$ & MAPE & MSE & {Maximum \\ error} \\
     \midrule
     Mean & 0.53 & 0.84 & 44.18\% & 0.33 & 2.12 \\    
     SD & $\pm0.08$ & $\pm0.06$ & $\pm4.69\%$ & $\pm0.05$ &  $\pm0.30$\\   
     \bottomrule
    \end{tblr}
    \caption{Performance metrics from the model.}
    \label{tab:Modelmetrics}
\end{table}

\begin{figure*}[t]
    \begin{center}
    \includegraphics[width=.9\textwidth]{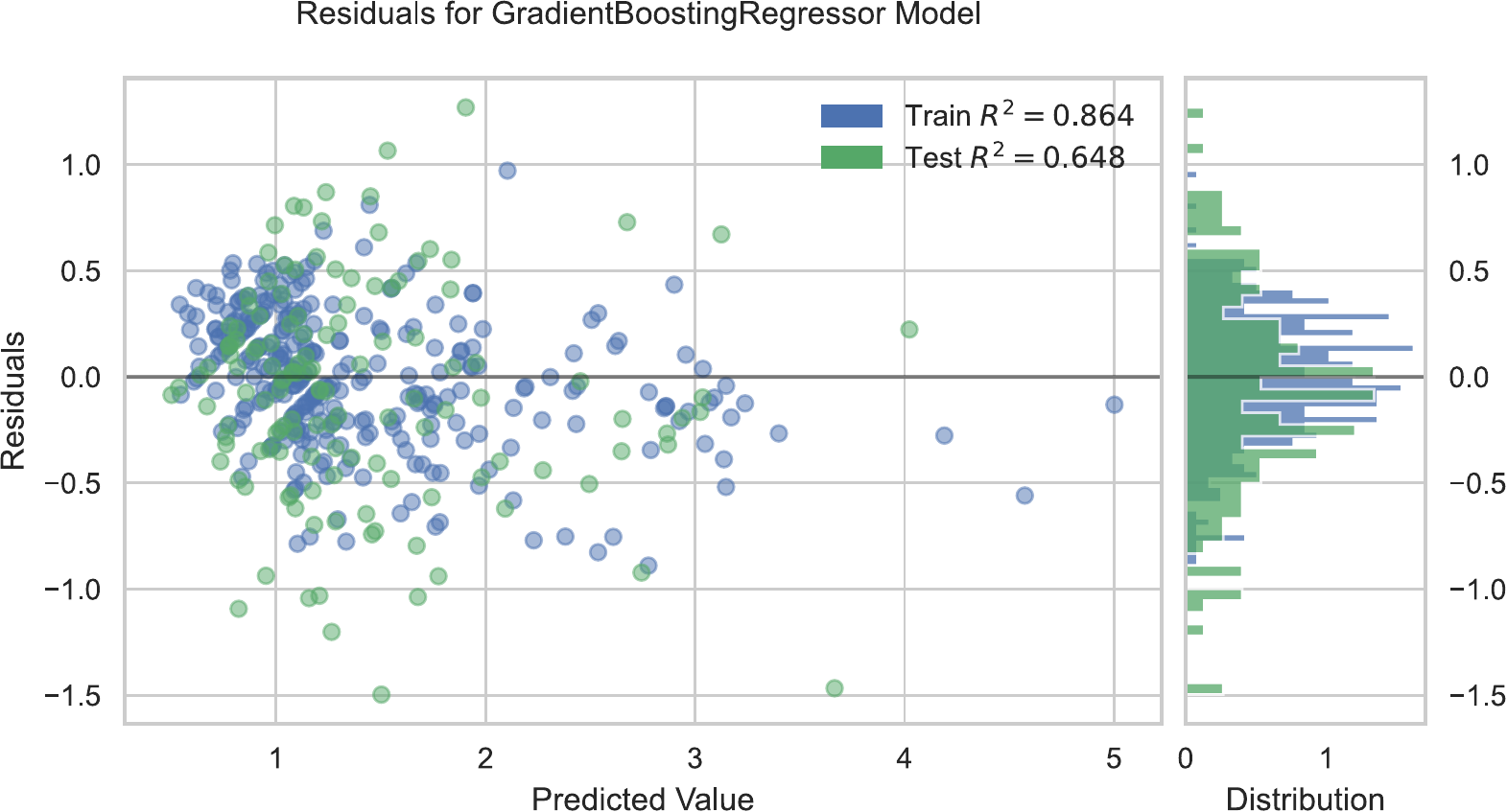}
    \end{center}
    \caption{Residuals of the model depicted using \citet{Bengfort2019Yellowbrick:Process} \textit{yellowbrick library} for one of the best outcomes of the model.}
    \label{fig:Residuals_Model}
\end{figure*}

Fig. \ref{fig:Predictions_Model} presents the predictions of the model classified for different lenses and compared with the real values. Fig. \ref{fig:Residuals_Model} shows the residuals of the model for the train and test dataset.

\begin{figure}
\begin{center}
    \includegraphics[width= \columnwidth]{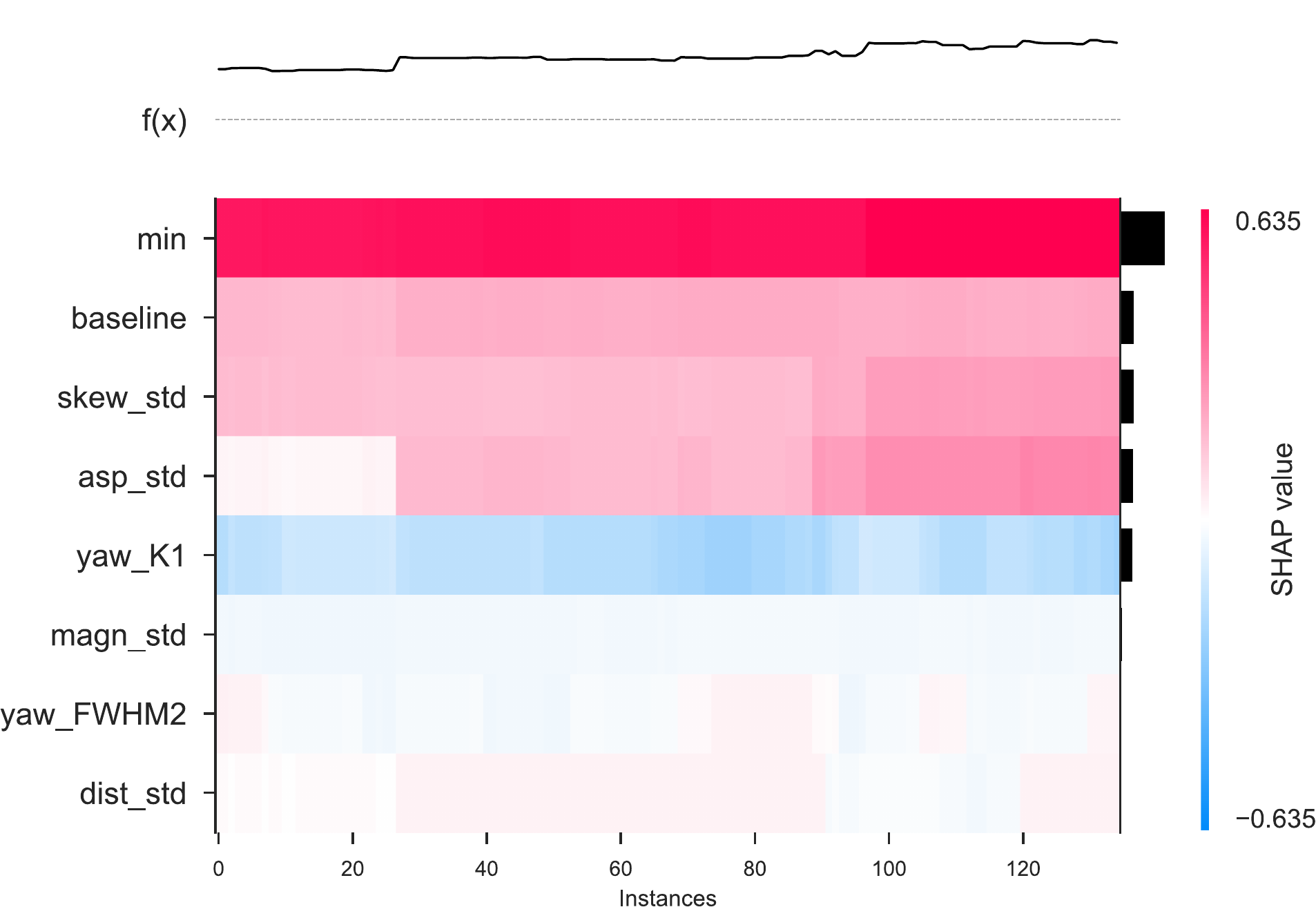}
\end{center}
\caption{SHAP values for the selected features of the model.}
\label{fig:SHAP_heatmap}
\end{figure}

The Fig. \ref{fig:SHAP_heatmap} presents the \textit{"SHAP"} values for the selected features of the model.

\section{Discussion} \label{Discussion}
This study shows that distortions of \acrshort{pals} can be simulated in \acrlong{VR} and that different lenses holding various degrees of distortions are perceived by the participants as having different levels of vexation, being those that presented a higher level of distortions experienced as more uncomfortable. In our case, the lens with spherical power \SI[retain-explicit-plus]{+2}{\diopter} and addition \SI[retain-explicit-plus]{+2}{\diopter} was rated the worst of all, followed by those without spherical power and addition \SI[retain-explicit-plus]{+3}{\diopter} and \SI[retain-explicit-plus]{+2}{\diopter}. 
One would expect the baseline (i.e. no distortions) to be rated as the most comfortable, but nevertheless, the lens distortions due to the negative spherical power and the +2 addition presented levels of discomfort equal to the baseline. A sum of several factors could have contributed to this result. On the one hand, these lens distortions have low levels of displacement in the central area, as can be seen in Fig. \ref{fig:3}, which could explain similar ratings to the baseline. 
We also cannot be entirely sure that the manufacturer has fully corrected in the software the distortions caused by the \acrshort{VR} headset lenses, and considering this headset uses a canted display, magnification from negative lenses may have made the image look more natural.
The subjective input scale was consistent given the Cronbach's index.

Although no correlation was found between the age of the subjects and the ratings given, it is important to mention that we have a reduced sample of young participants with a small variance. It can not be discarded that ratings might be different if an older group was tested, including presbyopes. Moreover, all of our subjects were emmetropes with an aim to avoid bias in our reduce sample, but usual ophthalmic wearers might like more the distortions that look alike to those they are habituated to.

Subjects' heading and gazing presented a significant variance across individuals, independently of the distortions applied, which is expected as it is known that these parameters are tied to each individual. Only one of the inverse power Batschelet distributions fitted for yaw varied significantly between the lenses and ratings, with a wider "+2/Add2" distribution, implying the more distortions, the less accurate our horizontal head movement is. The statistical results comparing against clusters of ratings should be taken with care. Although non-parametric tests were used, the number of measurements on the worse/higher ratings (4-5, after normalising) were quite low. For example, some parameters of yaw and roll were significantly different across rating groups, but no clear pattern was found, and no additional differences were found in the post-hoc analysis.

Regarding gaze movements, fewer fixations per minute lead to better ratings, this was not found between lens conditions, where there were no statistical differences, but +2/Add2 presented more fixations per minute. In this case, a pattern could be observed from the ratings, where the more fixations are made per minute, the more distress is perceived, and only in the 4-5 groups it was disrupted, probably due to the reduced sample of this groups. Thus, more distortions may have led to more fixations, and the more one fixates, the more one perceives those distortions, giving them worse ratings.
Furthermore, more saccades were performed in the textures with greater distortions, and ratings followed the same pattern but did not reach significance. Otherwise, the worse ratings presented shorter saccadic lengths and slower peak velocities in the saccades, but, in this case, nothing could be found for the different lens conditions. 
In general, gazing behaviour is highly tied to each individual, like the heading movements. However, more saccades and more fixations led to higher rates of discomfort, although that might have been coupled with the lens in use.

The lens condition dictated how much of the mean and standard deviation of displacement, magnification, rotation, differences in the aspect ratio, and skew were observed, a factor which influence the final rating. No differences across subjects were found, probably because even if they gaze differently, the area they cover and locations they gaze through were pretty similar.

\subsection*{Limitations of the study}
There are several limiting factors that may have constrained the study's outcome. One of the main hiccups of the data was the reduced amount of high/worse discomfort ratings (see Fig. \ref{fig:Residuals_Model}), precluding the drawing of some conclusions and probably limiting the model's accuracy. To gather more data with high ratings, the sample size must be increased by presenting more trials with higher distortions lenses, more lens conditions with a high amount of distortions, or simply measuring more subjects. In fact, including more lenses, with not only different prescription powers but also different designs may provide more information on individual characteristics and their influence on discomfort.

In this study, only "static distortions" were measured, i.e. the distortions of the \acrshort{pals} were only measured on the lens surface and presented as if looking through the main line of sight. Although the perceived distortions varied when looking through different areas, the distortions applied through the custom shader did not update with the gaze. Presenting "dynamic distortions" entails having previously measured and stored a texture of how each pixel shifts on the texture at each possible gaze point. Considering the \acrshort{fov} reported for this \acrshort{hmd}, one would need roughly ~18000 measurements per lens to do it on every degree of the visual field. This task becomes unfeasible unless a model is built that, from lesser measurements, interpolates the rest of the textures.

As already mentioned, the lenses used by the \acrshort{hmd} present several distortions that must be corrected through software. Therefore, the image presented on display is already warped to compensate for the lenses. It sets a limitation as only the manufacturer usually knows how well it is corrected. Additionally, this compensation only takes one fixed pupil position ('static distortions'), and only a recent paper \citep{Chan2022PredictingScenes} has started to look at how dynamic distortions affect the whole \acrshort{VR} experience and discomfort by inducing unintended optic flow.

Beyond distortions, progressive lenses present different blurring across the visual field, usually requiring a change in gaze behaviour to avoid blurry vision, which can additionally influence the comfort while wearing these types of lenses. In fact, the visual system is thought to be more tolerant to distortions than it is to blur \citep{Barbero2015GeometricalLenses}. Nevertheless, a future study should assess the discomfort that these lenses present due to blurring alone and perhaps a combination of both.

\subsection*{The model}
A gradient boosting regressor model was built. Although the model's accuracy on a single prediction is not outstanding, the model does behave as expected for the different lenses. It is capable of predicting ratings well within ranges defined for each lens condition. As mentioned in the limitations, the model could be further enhanced if more data was acquired, mostly to support the high discomfort ratings.

Although SHAP values \citep{Lundberg2017APredictions} were calculated to look for how model features act on predictions of discomfort, these values should be taken with care, especially when trying to understand decision tree models. The minimum rating given by the subject can be found as a main contributor. This could have occurred because the model may have learned to identify the subject from the minimum rating given, hence putting all individual preference weights into that decision, i.e. basing its decisions on its idea of which subject was estimating, even though this information was not. Other factors contributing to the predictions were the baseline values reported by the subject, which also could have served to acknowledge the participant, the SD of the observed skew, aspect ratio differences, magnification and a minor role of displacement. The concentration parameter of the first distribution fitted in yaw also contributed, which might be an indirect consequence of how straight was the head relative to a gazed target, upon a yaw movement, due to the different lenses.

It is not strange that skew was found to be a contributing factor to discomfort while "wearing" \acrshort{pals}. In fact, other studies already connected skew and adaptation \citep{Habtegiorgis2017AdaptationAftereffects,Rifai2020Motion-formScenes}. Magnification was also previously connected
to the so-called \textit{swim} effect, which relates to the illusory and variable seesaw-like movement of the visual field that originates with lateral head movements. This is further perceived with the nonuniform optical magnification effects between near and far objects \citep{Han2003DynamicEnvironment}. Magnification in the adaptation to spectacles was also mentioned to induce changes in the vestibular ocular reflex and discomfort \citep{Cannon1985TheVOR}, which might induce more discomfort in VR \citep{Chang2020VirtualMeasurements}.
\section{Conclusions} \label{Conclusions}
Similar to how using free-form technologies impacted the \acrshort{pals} market, reducing stock, costs, and expanding the fitting possibilities \citep{Alonso2019ModernOptics}, virtual reality holds the potential to become a testing bench for acceptance of new optical consumer solutions and thus to decrease the innovation costs of developing such lenses. \acrshort{VR} combines the naturality of performing a daily task with the ability to tweak specific conditions which might not be possible in the real world. 

This study proves it is possible to simulate, in virtual reality (up to a certain degree), how distortions affect our visual field given different progressive lenses. Besides the high inter-subject variability, every participant perceived different levels of discomfort for different amounts of distortions. Other traits were found beyond subjects' susceptibility or the distortions quantity defining the discomfort level that one can perceive with a specific lens in certain conditions, such as certain gazing or heading behaviours.

Finally, a machine learning model may help reduce the amount of testing required and help to understand what features contribute the most or why this sensation of discomfort does appear for some but not everyone. However, an ampler amount of data is required to obtain a more applicable model.
\section{Acknowledgements} \label{Acknowledgements}
The authors acknowledge support by the Open Access Publishing Fund of the University of Tübingen and the state of Baden-Württemberg through bwHPC.
\section{Declarations} \label{declarations}
\section*{Funding} This work was supported by the European Grant PLATYPUS (Grant Agreement No 734227), a Marie Sklodowska-Curie RISE initiative. Authors MGG \&  YS are employees of the University of Tübingen (E), SW is employed by Carl Zeiss Vision International GmbH (E) and is a scientist at the University Tübingen. TW is employed (E) by Western Sydney University. According to the journal policy, they declare their employment positions.

The founders did not play any additional role in the study design, data collection, and analysis, decision to publish, or preparation of the manuscript. The specific roles of these authors are articulated in the 'author contributions' statement.

\section*{Conflicts of interest/Competing interests}
The authors declare that the research was conducted in the absence of any commercial or financial relationships that could be construed as a potential conflict of interest.

\section*{Availability of data and material (data transparency)}
The datasets generated during and/or analysed during the current study are available from the corresponding author on reasonable request.

\section*{Authors' contributions} 
Conceptualization: MGG, TW \& SW. Formal analysis, methodology, software, validation and visualization: MGG \& YS. Data curation, investigation, writing-original draft: MGG. Writing-review and editing: MGG, YS, TW \& SW. Supervision, resources and project administration: TW \& SW. Funding acquisition: SW.

\section*{Ethics} \label{ethics}
The ethics authorisation to perform the measurements was granted by the Ethics Committee at the Medical Faculty of the Eberhard-Karls University and the University Hospital Tübingen with the ID 986/2020BO2.


\bibliographystyle{plainnat}

\Urlmuskip=0mu plus 1mu\relax

\bibliography{references}
\end{document}